\documentclass{article}

\usepackage{PRIMEarxiv}

\usepackage[utf8]{inputenc} % allow utf-8 input
\usepackage[T1]{fontenc}    % use 8-bit T1 fonts
\usepackage{hyperref}       % hyperlinks
\usepackage{url}            % simple URL typesetting
\usepackage{booktabs}       % professional-quality tables
\usepackage{amsfonts}       % blackboard math symbols
\usepackage{nicefrac}       % compact symbols for 1/2, etc.
\usepackage{microtype}      % microtypography
\usepackage{lipsum}
\usepackage{graphicx}
\graphicspath{{media/}}     % organize your images and other figures under media/ folder
\usepackage{amsmath,amssymb,amsfonts}
\usepackage{algorithmic}
\usepackage{graphicx}
\usepackage{textcomp}
\usepackage{xcolor}
\def\BibTeX{{\rm B\kern-.05em{\sc i\kern-.025em b}\kern-.08em
    T\kern-.1667em\lower.7ex\hbox{E}\kern-.125emX}}

\usepackage{amsmath,amssymb,amsfonts}
\usepackage{booktabs}
\usepackage{xspace}
\usepackage{multirow}
\usepackage{balance}
\usepackage{paralist}
\usepackage{microtype}
\usepackage{listings}
\usepackage{hyperref} 

\usepackage{xcolor}
\usepackage{subcaption}

\newcommand{\tool}{\textsc{Flow\-Repair}\@\xspace} %	MetAmorphic Runtime MOniToring

\AtBeginDocument{%
  \providecommand\BibTeX{{%
    \normalfont B\kern-0.5em{\scshape i\kern-0.25em b}\kern-0.8em\TeX}}}

\lstdefinestyle{javastyle}{
    backgroundcolor=\color{white},   
    commentstyle=\color{teal},
    keywordstyle=\color{purple},
    numberstyle=\tiny\color{gray},
    basicstyle=\ttfamily\scriptsize,
    breakatwhitespace=false,         
    breaklines=true,                 
    captionpos=b,                    
    keepspaces=true,                 
    numbers=left,                    
    numbersep=5pt,                  
    showspaces=false,                
    showstringspaces=false,
    showtabs=false,                  
    tabsize=1
}
\usepackage{float}
\usepackage[ruled, vlined, linesnumbered]{algorithm2e}
\SetAlFnt{\small}
\SetAlCapNameFnt{\small}
\SetAlCapFnt{\small}
\SetFuncSty{textsc}
\AtBeginEnvironment{algorithm}{
	\DontPrintSemicolon
}

\newsavebox\CBox

\RequirePackage{float}
\RequirePackage[skins]{tcolorbox}
\newtcolorbox{custombox}[1]{
	colback=gray!10,
	colframe=gray!70,
	left=1mm,
	right=1mm,
	top=1mm,
	bottom=1mm,
	fonttitle=\bfseries,
	arc=0mm,
	leftrule=1mm,
	rightrule=0mm,
	toprule=0mm,
	bottomrule=0mm,
	notitle,
	before=\par\smallskip\noindent,
	before upper={\textbf{#1: } },
}
\newtcolorbox{probdefinition}[1]{
	colback=gray!10,
	colframe=gray!22,
	left=1mm,
	right=1mm,
	top=1mm,
	bottom=1mm,
	fonttitle=\bfseries,
	arc=2mm,
	leftrule=0mm,
	rightrule=1mm,
	toprule=0mm,
	bottomrule=1mm,
	notitle,
	before=\par\smallskip\noindent,
	before upper={\textbf{#1: } },
}

\title{Search-based Automated Program Repair of CPS Controllers Modeled in Simulink-Stateflow}

\author{
  Aitor Arrieta \\
  Mondragon University \\
  Mondragon, Spain \\
  \texttt{aarrieta@mondragon.edu}  \\
  \And
  Pablo Valle \\
  Mondragon University \\
  Mondragon, Spain \\
  \texttt{pvalle@mondragon.edu}  \\
   \And
  Shaukat Ali \\
  Simula Research Laboratory \\
  Oslo Metropolitan University\\
  Oslo, Norway\\
  \texttt{shaukat@simula.no} \\
}

\begin{document}
\maketitle

\begin{abstract}
Stateflow models are widely used in the industry to model the high-level control logic of Cyber-Physical Systems (CPSs) in Simulink--the defacto CPS simulator. Many approaches exist to test Simulink models, but once a fault is detected, the process to repair it remains manual. Such a manual process increases the software development cost, making it paramount to develop novel techniques that reduce this cost. Automated Program Repair (APR) techniques can significantly reduce the time for fixing bugs by automatically generating patches. However, current approaches face scalability issues to be applicable in the CPS context. To deal with this problem, we propose an automated search-based approach called \tool, explicitly designed to repair Stateflow models. The novelty of \tool includes, (1) a new algorithm that combines global and local search for patch generation; (2) a definition of novel repair objectives (e.g., the time a fault remained active) specifically designed for repairing CPSs; and (3) a set of mutation operators to repair Stateflow models automatically. We evaluated \tool with three different case study systems and a total of nine faulty stateflow models. Our experiments suggest that (1) \tool can fix bugs in stateflow models, including models with multiple faults; (2) \tool surpasses or performs similarly to a baseline APR technique inspired by a well-known CPS program repair approach. Besides, we provide both a replication package and a live repository, paving the way towards the APR of CPSs modeled in Simulink.%  
\end{abstract}

% keywords can be removed
\keywords{Automated Program Repair \and Cyber-Physical Systems \and Simulink \and Stateflow}

\section{Introduction}

Cyber-Physical Systems (CPSs) integrate digital cyber computations with physical processes \cite{derler2011modeling}. As these systems involve complex and expensive hardware, their software is usually tested through simulation-based testing in early stages of development~\cite{matinnejad2016automated,matinnejad2018test,arrieta2019search,khatiri2023simulation,ben2016testing,abdessalem2018testing}. MATLAB/Simulink has been the de-facto CPS modeling and simulation tool~\cite{matinnejad2016automated,matinnejad2018test}, allowing complex features like automated test generation and standard-compliant code generation. The control of CPSs is usually divided into high-level decision-making components and low-level controllers~\cite{mandrioli2023stress}. The high-level decision-making controllers are typically delegated to Stateflow models when the CPS is modeled in Simulink. Stateflow is a graphical language for modeling state charts. It is common to find logic bugs in Simulink models involving Stateflow models. While many approaches have been focusing on automated techniques to generate test cases for Simulink ~\cite{matinnejad2016automated,matinnejad2018test,arrieta2017employing,menghi2020approximation,menghi2019generating}, once a fault has been detected, the repair process has remained manual. This poses 
a high cost to engineers, and therefore, it is necessary to develop novel techniques for the automated program repair (APR) of Simulink models.

Recently, the area of APR has significantly grown, proposing novel techniques to automatically repair software bugs in different languages (e.g., Java, C, Python). These approaches encompass different techniques, such as search-based~\cite{le2011genprog,yuan2018arja}, semantic-based~\cite{nguyen2013semfix}, and neural-machine based repair approaches~\cite{tufano2018empirical}. More recently, the APR community has focused on using Large Language Models (LLMs)~\cite{xia2023automated,xia2022less} and conversational techniques (i.e., ChatGPT)~\cite{SobaniaChatGPT2023} for APR, with impressive results. However, repairing CPSs' software has two core challenges that limit the applicability of existing APR approaches to CPSs:

    \noindent\textbf{Challenge 1 -- Long test execution time:} When testing CPSs, test cases need to be executed at the system level employing simulation-based techniques~\cite{ValleSEIP2023,abdessalem2020automated}. The physical system and the software control need to be integrated because the physical outcomes affect the software and vice-versa~\cite{abdessalem2020automated}. This is commonly known as online testing~\cite{haq2020comparing,haq2021can}. These physical systems are usually modeled through complex mathematical models that are computationally expensive, and often require rendering 3D images, which increases test execution time~\cite{haq2020comparing,haq2021can}. For instance, previous APR approaches targeting industrial CPSs report average test execution times of around 5 minutes~\cite{ValleSEIP2023} and 20 to 30 minutes~\cite{abdessalem2020automated}, for different industrial case study systems. Therefore, unlike most APR approaches, which employ unit test cases that are executed in milliseconds, in the context of CPSs, the execution time poses limitations on the sizes of test suites used for repair~\cite{abdessalem2020automated}.

    \noindent\textbf{Challenge 2 -- Inadequate repair objectives:}  Most existing repair objectives focus on high-level passing and failing of test cases to guide the search~\cite{goues2019automated}. The problem with CPSs is that many test cases cannot be executed due to the aforementioned challenge. On the other hand, it is common that test cases for CPSs are generated with a falsification-based approach (e.g.,~\cite{menghi2020approximation,nejati2019evaluating}). This means that when a system requirement is violated, the test generation process stops, returning a single failure-inducing test input. This limits the guidance of search-based APR approaches, converting the search process into a random one due to the flat fitness landscape. As a result, novel repair objectives need to be investigated.

To address the above-mentioned challenges, we instead, propose a search-based APR method for Stateflow models, paving the way towards the APR of CPSs. Specifically, we solve the first issue by proposing an iterative algorithm that handles an archive of partial patches instead of a population-based search algorithm. The key idea behind the algorithm is to use a global search algorithm to explore different kinds of patches at different parts of the code. When a partial patch is found (i.e., a patch that enhances the buggy code but does not fix it entirely), the global search is swapped by a local search process that exploits patch generation in a restricted Stateflow area. As a result, we solve the long execution time of the population-based search algorithms. The second issue is solved by redefining the repair objectives. Specifically, instead of focusing on the number of pass and fail test cases to guide the search, we define three novel repair objective functions that consider CPS particularities, such as the notions of time and failure severity. In summary, our paper makes the following contributions:

\begin{itemize}
    \item \textbf{Method: }We propose a novel method to repair faults in Stateflow models that control CPSs. This method encompasses a novel search algorithm, which integrates global and local search and adopts novel repair mutations for Stateflow models. In addition, we propose three new repair objectives that are generic for any kind of APR tool for CPSs. To the best of our knowledge, this is the first paper that targets the APR of Simulink models, the de-facto CPS modeling and simulation tool~\cite{matinnejad2016automated,matinnejad2018test}.
    \item \textbf{Tool: }We prototype the method in a tool and provide it as open-source on GitHub, together with a replication package on Zenodo. Our code and up-to-date progress can be found at Github. See the replication package for details.
    \item \textbf{Evaluation and Dataset: }We carry out a comprehensive evaluation of our tool by complementing an existing dataset of Stateflow bugs~\cite{ayesh2022two} with a new dataset encompassing two new CPS models. In total, the dataset encompasses 9 real bugs, out of which 7 are provided by this paper. Despite not being a large number, in terms of the number of faults, it is the largest evaluation related to APR in the field of CPSs.
\end{itemize}

The rest of the paper is structured as follows: Section \ref{sec:background} presents relevant background, and Section \ref{sec:approach} introduces our approach. Empirical evaluation, results, and threats are presented in Section \ref{sec:evaluation}, Section \ref{sec:discussion}, and Section \ref{sec:threats}, respectively. We position our work with respect to relevant works in Section \ref{sec:relatedwork}. We conclude the paper in Section \ref{sec:conclusion}.

%\aitor{1) Mutation testing, what it is, why it is important. 2) mutation testing of DNN (different works). 3) DeepCrime, what it is, and why it is good to use this one. }

%\aitor{1) Metrics based on the MSE. However, the MSE is applied in offline mode. In contrast, physical testing is on-line, which means that a (wrong) decision taken by a DNN at a point of time $t$, can be corrected afterwards.}

%Testing citation \cite{humbatova2021deepcrime}.

%In this work we evaluate the behavior of 20 different mutants obtained from four different mutation operators proposed by Humbatova et al. \cite{humbatova2021deepcrime} in a physical robot. 

\section{Background} \label{sec:background}
\label{sec:background}

\subsection{Automated Program Repair}

 Given a buggy program $P$, the corresponding specification $S$ that makes $P$ fail and the program transformation operators $O$, Automated Program Repair (APR) can be formalized as a function $APR(P,S,O)$. $PT$ is the set of its all possible program variants by enumerating all operators $O$ on $P$. The problem of APR is to find a program variant $P’$ ($P\in PT$) that satisfies $S$. The difference in the software program between $P’$ and $P$ is known as a \textit{plausible patch}. $S$ is usually defined as \textit{passing} and \textit{failing} test cases, where passing test cases demonstrate the functionality that needs to be preserved, whereas the failing one demonstrates the bug~\cite{goues2019automated}. A plausible patch might be overfitted to $S$, therefore, an additional manual process may be required to validate that the plausible patch is a valid patch that fixes the bug.

 Three main techniques exist for APR. Search-based program repair techniques attempt to recast program repair to that of an optimization problem by navigating through the search space of program edits. The goal is usually to reduce the number of failing test cases while increasing the number of passing ones~\cite{le2011genprog}, although other objectives exist too (e.g., reduce the number of program edits~\cite{yuan2018arja}). Another approach involves semantic program repair~\cite{nguyen2013semfix}, which focuses on making the repair process explicit by deriving a specification from it. Starting with a correctness requirement, a constraint is deduced that describes the necessary changes. Subsequently, a minimal patch is generated to satisfy this constraint. The last technique is based on deep learning techniques. This is based on learning repair strategies from human patches. The learning-based repair techniques (e.g., Getafix~\cite{bader2019getafix}) first mine human patches that fix defects in existing software repositories, train a general repair rule and apply the model to buggy programs to produce patches. Besides, the use of LLMs has been recently leveraged for APR~\cite{xia2022less}. These, instead of mining and training an algorithm, make use of pre-trained LLMs. In this paper, we focus on the first type of technique, i.e., search-based program repair.
 
\subsection{Stateflow Models}
Stateflow \cite{stateflowMATLAB} is a visual programming language and simulation environment for Simulink created by MathWorks. It is primarily used for modeling and simulating complex control systems and state-based behaviors in various applications. The main components of a Stateflow diagram are the states and the transitions. The states are a set of events shown as a box representing a set of behaviors or actions that the system can perform at that moment. To switch from one state to another, a transition must be triggered. The transitions are represented as arrows that connect the states. To activate a transition, a set of boolean conditions must be met.

\begin{figure}[h!]
    \centering
    \includegraphics[width=0.9\linewidth]{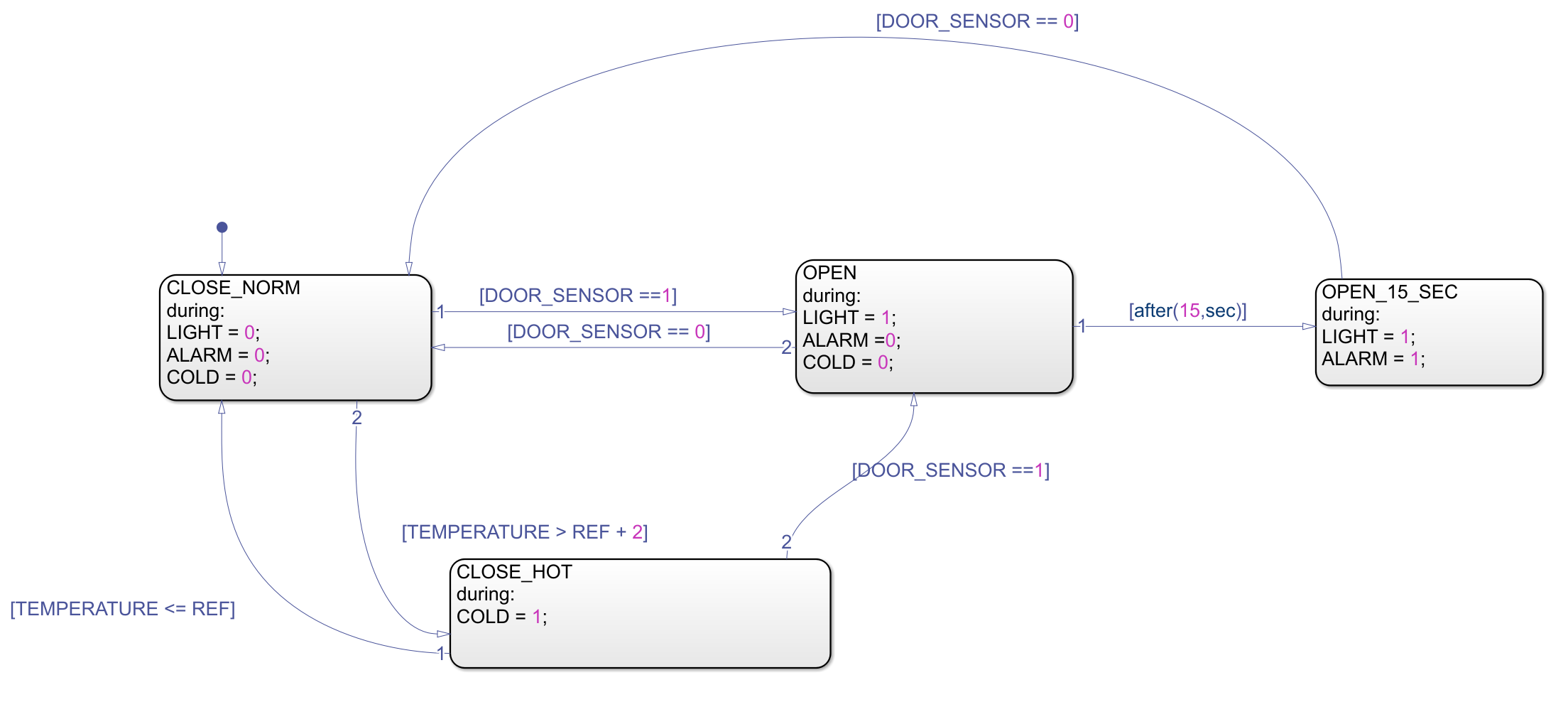}
    \caption{Stateflow diagram of a fridge temperature controller}
    \label{fig:stateflow}
\end{figure}

Figure~\ref{fig:stateflow} shows an example of a Stateflow diagram that controls the temperature of a fridge. The example has four states: (1) CLOSE\_NORM indicates that the door of the system is closed and the temperature is cold enough not to activate the cold-down function; (2) CLOSE\_HOT indicates that the temperature is too warm, and needs to be cold down through the variable COLD; if the door is opened, the system enters in the state (3) OPEN. When this happens, the cold-down function is disabled, and a light is turned on through the variable LIGHT. If the door remains open for 15 seconds, it enters the state (4) OPEN\_15\_SEC. This state triggers an alarm to indicate to the user that the door shall be closed, which is carried out through the the ALARM variable.

\section{FlowRepair} \label{sec:approach}
\label{sec:approach}

\subsection{Overview of the APR Process}

\begin{figure*}[th!]
    \centering
    \includegraphics[width=.95\linewidth, trim = 0 30 0 0]{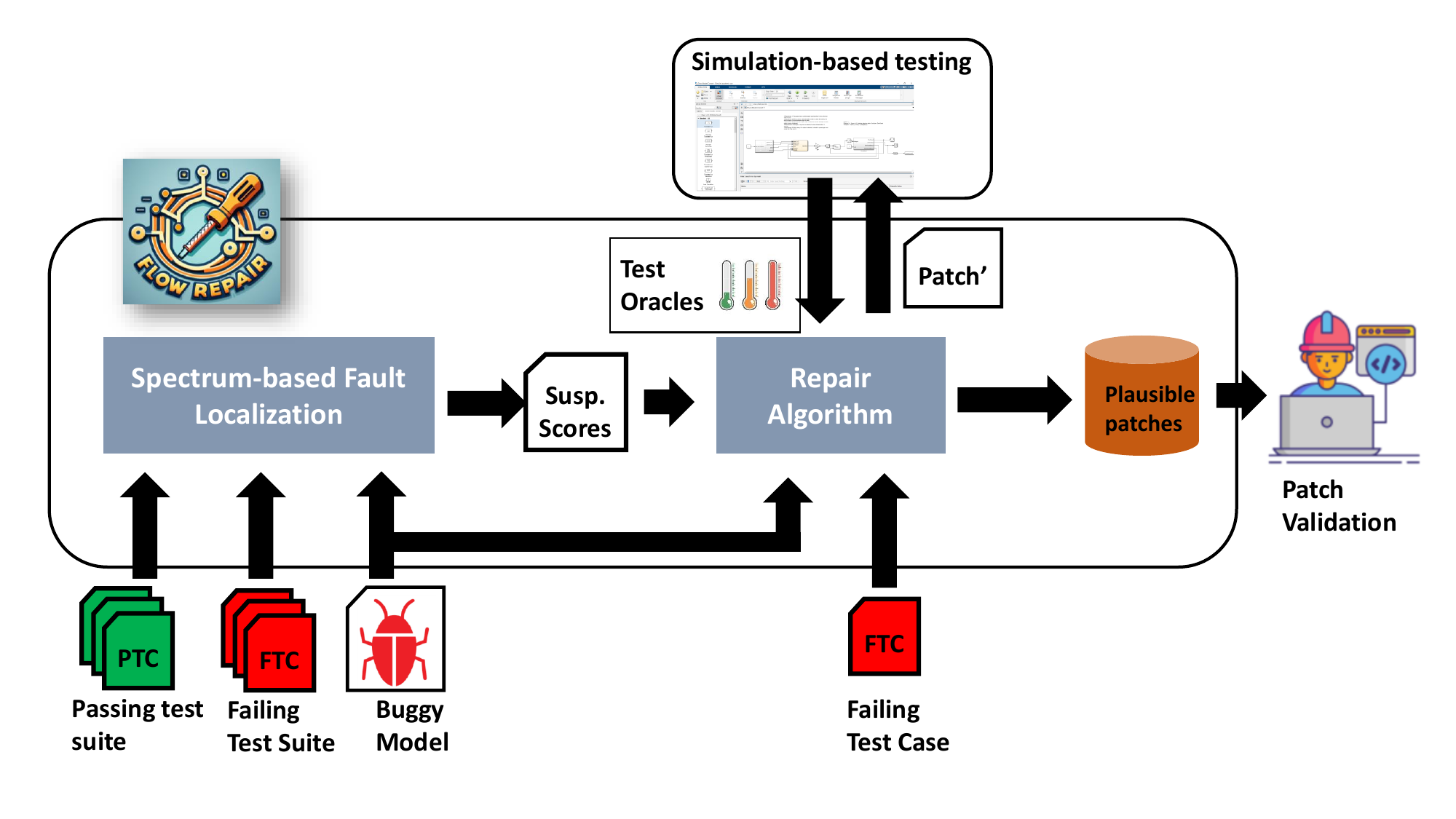}
    \caption{High-level overview of \tool}
    \label{fig:Approach}
\end{figure*}

Figure \ref{fig:Approach} shows an overview of \tool. Like most APR approaches, \tool's first stage involves a fault localization process. This narrows the search space for finding plausible patches, making the repair process more efficient. We opted for using Spectrum-based Fault Localization (SBFL) for this task, as it is one of the most commonly employed approaches to localize faults~\cite{wong2016survey}. To this end, in the first stage, the tool automatically instruments the Stateflow models inside a Simulink model. This is carried out to obtain, for each test case, which of the transitions and states were executed. We also considered employing Simulink Design Verifier (SDV) for this task. SDV  is a Mathwork's toolbox that gives the option for analyzing coverage data. However, we discarded this option because an additional license is required, restricting \tool's access to any practitioner and researcher without the license. When the model is instrumented, \tool executes the entire test suite. By tracing the data of each failing and passing test case with the executed transitions and states, we obtain, for each transition and state of the model, a suspiciousness score. The suspiciousness scores of the components of the stateflow models in Simulink models are extracted by using the well-known SBFL Tarantula metric~\cite{wong2016survey}. Note, however, that the tool is flexible and that any other SBFL metric can be easily incorporated. The suspiciousness scores are employed in the upcoming repairing stage.

After obtaining the suspiciousness scores of each transition and state in the Simulink model, \tool triggers the repair algorithm, explained in Section \ref{sec:repairAlgorithm}. This algorithm combines a global search process with a local one to propose patches. These patches are executed by using simulation-based testing, which returns simulation results, which are processed by \tool to obtain a set of repair objectives (explained in Section \ref{sec:objectives}). Notably, a single failing test case is employed during the repair process. This is (1) because executing simulation-based test cases of CPSs is highly expensive, thus, increasing the tool's efficiency; and (2) because CPSs are usually tested by employing falsification-based approaches that stop generating test cases once a failure is triggered~\cite{menghi2020approximation}, therefore, having a single failing test case in the test suite. The repair algorithm stops when a time budget is exceeded. All the plausible patches are given to the software engineer for manual inspection and validation.

\subsection{Repair Algorithm}
\label{sec:repairAlgorithm}

Algorithm \ref{alg:FlowRepair} shows the overview of \tool's steps for repairing Stateflow bugs. The algorithm takes as inputs a Simulink model to repair (modelToRepair) and the suspiciousness ranking obtained with SBFL (suspiciousnessRanking). As output, it provides an archive of plausible patches, i.e., patches that pass the original test suite and that should undergo a manual validation process to assess their correctness. The algorithm maintains two archives: \textit{PlausiblePatchArch}, containing all plausible patches found during the repair process, and \textit{PartialArchive} having partial patches found during the repair process. A partial patch is a partially repaired solution, i.e., a solution that enhances the previous model, but that still does not pass the failing test case. These two archives are initialized in Lines 1 and 2 of Algorithm~\ref{alg:FlowRepair}.

\begin{algorithm*}[ht]
\caption{Overview of \tool's repair algorithm}\label{alg:FlowRepair}
   % \begin{algorithmic}
    \KwIn{modelToRepair \\ suspiciousnessRanking}
    \KwOut{PlausiblePatchArch}
    PlausiblePatchArch $\gets$ $\varnothing$ \\
    PartialArchive $\gets$ FaultyModel \\
    \While{TimeBudgetNotExceeded}{
        VerdictEnhanced $\gets$ False\\
        \While{VerdictEnhanced == False and TimeBudgetNotExceeded}{
            modelToRepair $\gets$ \textsc{selectModel}(PartialArchive)\\
            mutatedModel, mutatedComponent, mutationOperator $\gets$ \textsc{applyGlobalMutations}(modelToRepair,suspiciousnessRanking)\\
            verdict $\gets$ \textsc{executeTests}(mutatedModel)\\
            \uIf{verdict == pass}{
                 PlausiblePatchArch $\gets$ PlausiblePatchArch $\cup$ mutatedModel \\
            }\ElseIf{\textsc{checkVerdictEnhanced}(verdict,Archive)}{
                VerdictEnhanced $\gets$ true;\\
                PartialArchive $\gets$ PartialArchive $\cup$ mutatedModel \\
                modelToApplyLocalMutations $\gets$ mutatedModel \\
            }        
        }
        numOfLocalTries $\gets$ 0\\
        \While{numOfLocalTries $<$ totalLocalTries and TimeBudgetNotExceeded}{
            numOfLocalTries $\gets$ numOfLocalTries + 1 \\
            locallyMutatedModel $\gets$ \textsc{applyLocalMutations}(modelToApplyLocalMutations,mutatedComponent, mutationOperator) \\
            verdict $\gets$ \textsc{executeTests}(locallyMutatedModel)\\
            \uIf{verdict == pass}{
                    PlausiblePatchArch $\gets$ PlausiblePatchArch $\cup$ locallyMutatedModel \\
                }\ElseIf{\textsc{checkVerdictEnhanced}(verdict,Archive)}{
                    PartialArchive $\gets$ PartialArchive $\cup$ locallyMutatedModel \\
                    modelToApplyLocalMutations $\gets$ locallyMutatedModel \\
                    numOfLocalTries $\gets$ 0
                }   
        }
        \textsc{ReorganizeArchive}(PartialArchive)
    }
\end{algorithm*}

After this, the algorithm enters in a while loop (Lines 3-26, Algorithm~\ref{alg:FlowRepair}) that finishes when the time budget has been exceeded. This algorithm has two main parts, divided into two while loops. On the one hand, the first part of \tool, lines 5-14 in Algorithm~\ref{alg:FlowRepair}, applies a global search routine to find plausible patches or partial patches. To that end, the algorithm first selects the model to be repaired among all models included in PartialArchive. This selection is carried out randomly. After, this model is mutated (Line 6, Algorithm~\ref{alg:FlowRepair}) by applying a mutation to the Stateflow model of the CPS based on the developed mutation operators (Section \ref{sec:mutations}). To select the item of the model to be mutated (i.e., a state or a transition of the model), the algorithm uses the suspiciousness score information of each state/transition of the Stateflow model. Specifically, we employ a roulette wheel selection, which gives a higher probability of being mutated to those states and transitions with higher suspiciousness scores, whereas those with the lowest suspiciousness scores have lower chances of being selected. Once this is mutated, the model is tested (Line 7, Algorithm~\ref{alg:FlowRepair}). The test execution provides a verdict. If the verdict is a pass, we include the mutated model in the archive of plausible patches (Line 9, Algorithm~\ref{alg:FlowRepair}). If the verdict is not a pass, we check whether the verdict was enhanced with respect to the selected model (Line 11, Algorithm~\ref{alg:FlowRepair}). We consider a verdict is enhanced if any of the repair objectives is enhanced and none of the repair objectives is worsened. If the verdict is enhanced, then we include the mutated model in the archive of partial patches (Line 13, Algorithm~\ref{alg:FlowRepair}), and we trigger the local repair process.

On the other hand, the second part of \tool's repair algorithm applies a local search routine to locally enhance a partial patch (Lines 16-25, Algorithm~\ref{alg:FlowRepair}). This part of the algorithm is triggered when the global search routine finds a partial patch (i.e., an enhancement based on the repair objectives, explained in Section~\ref{sec:objectives}). The local search process exploits the search of patches by applying local mutations in the element of the Stateflow model (i.e., state or transition) that was mutated during the global search routine and enhanced the verdict. We limit the number of local tries to a predefined number (in our experiments to 30). We limit this number due to two main reasons. Firstly, because the enhancement could be due to an overfitted partial patch. Secondly, to avoid the search algorithm getting trapped in a local optima, which is common in local search processes. After applying the local mutation (Line 18, Algorithm~\ref{alg:FlowRepair}), we execute the failing test case to the model (Line 19, Algorithm~\ref{alg:FlowRepair}). If the verdict is a pass, then we include the model in the archive of plausible patches (Line 21, Algorithm~\ref{alg:FlowRepair}), and continue with the local mutation process to find additional plausible patches. If a partial patch is found, i.e., the verdict is enhanced but not fully passed, then a partial patch is included in the archive of partial patches, and the local search process is restarted with the new partial patch model (Line 24, Algorithm~\ref{alg:FlowRepair}) and initializing the number of local tries to 0 (Line 25, Algorithm~\ref{alg:FlowRepair}).

\begin{figure*}[th!]
    \centering
    \includegraphics[width=1\linewidth, trim = 0 320 0 0]{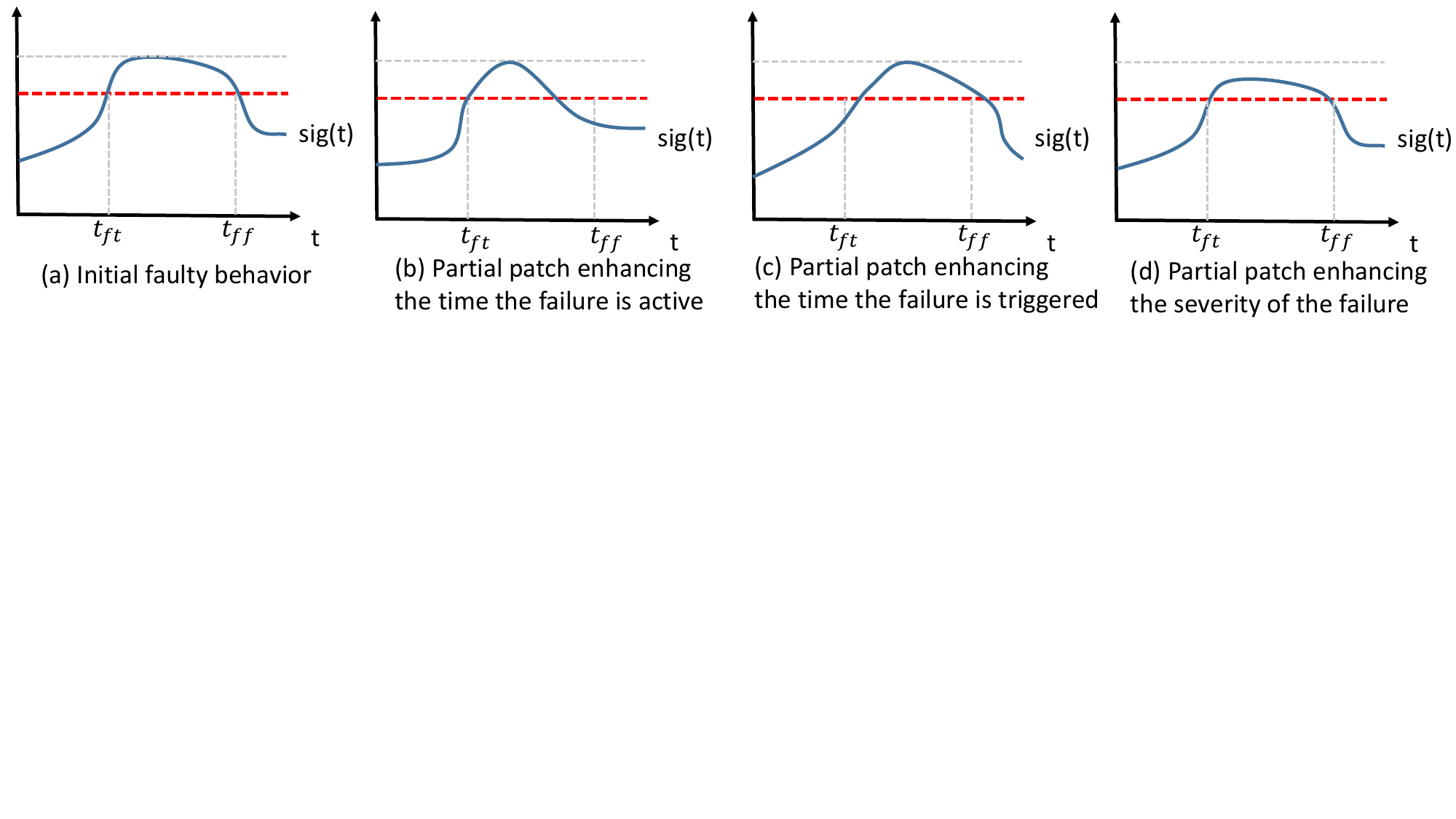}
    \caption{Given a signal over time ($sig(t)$) that violates the requirement involving the exceed of the red signal, an example of three repair objectives being improved }
    \label{fig:repairObjectives}
\end{figure*}

The local search process is finished either when the total time budget has been exceeded or when the number of local tries has exceeded the predefined number of total tries. When finishing the local search process, the archive of partial patches is reorganized. This is because it may include many partial patches, making the global search exhaustive. This reorganization can be done by following different strategies, such as considering Pareto-optimality among the selected repair objectives. After carrying out a set of preliminary experiments, we saw that this archive did not exceed a large number of partial patches; thus, we decided not to remove any partial patch from the archive in the current version of \tool. Moreover, we found that some partial patches may be overfitted (i.e., not correct). Therefore, removing patches from this partial patch may pose the risk of removing non-overfitted partial patches, which would increase the probability of not producing valid patches. Therefore, unlike previous APR approaches (e.g.,~\cite{abdessalem2020automated,ValleSEIP2023}), we found it sensible not to remove patches from this archive.

\subsection{Repair Objectives}
\label{sec:objectives}
Search-based APR approaches require a set of repair objectives to guide the search. Most approaches (e.g., GenProg~\cite{le2011genprog}, Arja~\cite{yuan2018arja}) employ the number of pass and failed test cases for guidance. However, in the context of CPSs, as explained in previous sections, executing a long number of test cases is impossible. Moreover, it is typical to only have a single failure-inducing test case. Subsequently, current repair objectives that are based on the number of failed and passed test cases are invalid for search-based APR in the context of CPSs, as the fitness landscape would become flat, precluding an appropriate search process.

To solve this issue, we take advantage of the notion of time and the severity of failures for guiding the search algorithm, provided that test cases in the context of CPSs are signals over time. 

\textbf{Motivating Example: }Consider as an example Figure \ref{fig:repairObjectives}, where the output signal over time ($sig(t)$) should not exceed the threshold value delimited by the red dashed line. Note, however, that in Figure \ref{fig:repairObjectives}-a this requirement is violated. The time at which the signal violates the requirement begins at time $t_{ft}$, and the time at which the failure finishes is $t_{ff}$.

Specifically, we define three repair objectives, out of which two are novel for the field of APR:

\begin{itemize}
    \item \textbf{Time the failure is active: }When executing the simulation of the CPS with a failure-inducing test case, we measure the time a failure is active. That is when the CPS is considered not to be working properly. The goal of \tool is to reduce this metric, as we conjecture that the lower the time a failure is active, the closer the repair algorithm is to providing a plausible patch. An example of this repair objective being enhanced by a partial patch is shown in Figure \ref{fig:repairObjectives}-b. In this case, it can be observed that the time the failure is active is less than $t_{ff}-t_{ft}$.
    \item \textbf{Time the failure is triggered: }When executing the simulation of the CPS with a failure-inducing test case, we measure when the failure is triggered. If, for a potential patch, the time at which the failure is triggered is postponed with respect to the selected buggy model from the archive, we consider that the potential patch is closer to being a plausible patch. Therefore, the goal of \tool is to increase the time at which the failure is triggered. An example of this repair objective being enhanced by a partial patch is shown in Figure \ref{fig:repairObjectives}-c. In this case, it can be observed that the time the failure is triggered is higher than $t_{ft}$. This repair objective is particularly interesting in those cases where the CPS does not recover from the failure (e.g., in autonomous driving systems, it is common that the vehicle goes out of bounds).
    \item \textbf{Severity of the failure: }Test verdicts provided by test oracles in the context of CPSs are not only boolean classifications of either pass or fail, but also provide, in the case of failed test cases, the severity of the failure, and in the case of pass test cases how far they are from their optimal performance~\cite{menghi2019generating}. As a result, most APR techniques applied in the field of CPSs have relied on the severity of the failure to guide the search process towards finding plausible patches~\cite{ValleSEIP2023,abdessalem2018testing,lyu2023autorepair}. Since this objective can help repair the Stateflow model too, and is complementary to the two new novel repair objective functions we define, we also include it in \tool. An example of this repair objective being enhanced by a partial patch is shown in Figure \ref{fig:repairObjectives}-d. In this case, it can be observed that although $sig(t)$ is still above the established threshold, its maximum value does not reach the grey dashed line. 

\end{itemize}

\subsection{Mutation Operators}
\label{sec:mutations}

The mutation operators of \tool represent program edits on the Stateflow model. Similar to other search-based APR tools, like GenProg~\cite{le2011genprog} or Arja~\cite{yuan2018arja}, \tool provides three types of mutation operations in the faulty Stateflow model: \textit{delete}, \textit{replace}, and \textit{insert}. More specifically, for suspicious components of the Stateflow model, \tool applies deletion (D), replacement (R), or insertion (I) patches to repair the buggy component(s). The considered components of the Stateflow model apply either to States or Transitions of the model. A problem with search-based APR tools is the number of implemented mutation operators. On the one hand, the lower the number of mutation operators, the lower the probability of generating a correct patch. On the other hand, the higher the number of operators, the larger it is the search space, which could impact the scalability of the approach. \tool contains a total of 15 mutation operators, which is large enough to provide sufficiently high repairability to generate correct patches in many different buggy scenarios. Besides, we employ these mutation operators in two ways to solve the scalability problem that may suppose having many operators: applying global mutation operators and applying local operators (further explained in Section \ref{sec:glolocalmutations}). We now explain the 15 selected operators:

\begin{itemize}
    \item \textbf{Relational Operator Replacement (R):} It replaces a relational operator with another one (e.g., $>$ for $>=$). This operator can be applied to transitions. 
    \item \textbf{Conditional Operator Replacement (R):} It replaces a conditional operator with another one (e.g., $\&\&$ for $||$). This operator can be applied to transitions.
    \item \textbf{Mathematical Operator Replacement (R):} It replaces a mathematical operator for another one  (e.g., $+$ for $-$). This operator can be applied to both transitions and states.
    \item \textbf{Unit Change in After Function (R):} It changes the time unit in an after function (e.g., \textit{after(5,sec)} for \textit{after(5,msec)}). This operator can be applied to transitions.
    \item \textbf{Numerical Replacement Operator (R):} It changes a numeric value for another value (e.g., 1 for -1). This operator can be applied to both transitions and states.
    \item \textbf{Transition Destination Replacement (R):} It changes the destination state of a transition to another state; the destination state is randomly chosen.
    \item \textbf{Transition Root Replacement (R):} It changes the origin state of a transition to another state; the new origin state is randomly chosen.
    \item \textbf{Initial Transition Change (R):} It changes the initial transition; consequently, it can only be applied in those initial transitions.
    \item \textbf{State Deletion (D):} It deletes one of the states of a Stateflow model. It can be applied to states, but all affected transitions are also removed.
    \item \textbf{Transition Deletion (D):} It deletes one of the transitions of the Stateflow model. This operator can be applied to transitions.
    \item \textbf{State Variable Deletion (D):} It deletes a variable from a state. This operator can be applied to states.
    \item \textbf{Transition Condition Deletion (D):} It deletes a condition from a transition. This operator can only be applied to those transitions with multiple conditions.
    \item \textbf{Mathematical Operation Insertion (I):} It inserts a mathematical operation based on the available inputs and outputs of the Stateflow model in a targeted transition. This operator can be applied to states.
    \item \textbf{Variable Insertion (I):} It inserts a new variable in a state. This operator can be applied to states.
    \item \textbf{Condition Insertion (I):} It inserts a new condition in a transition; thus, this operator can be applied to transitions.

\end{itemize}

\subsection{Global vs Local Mutations}
\label{sec:glolocalmutations}

To generate a patch, we apply a mutation to a Stateflow model based on one of the 15 mutation operators defined in Section~\ref{sec:mutations}. However, we do this in two ways, depending on the search routine at which the operator is triggered. We define these routines as global or local search. During \textit{global search}, the goal is to explore a large set of patches across different (suspicious) components of the Stateflow model. To generate a patch during the global search, the global mutation operator obtains one model to be repaired and the suspiciousness ranking for all the components in such a model (Line 7, Algorithm~\ref{alg:FlowRepair}). A component (either state or transition) of the Stateflow model is selected to be mutated. This selection is carried out via a roulette-wheel selection, which gives a higher mutation probability to a component with a higher suspiciousness and a lower probability to a component with a lower suspiciousness score. Based on this, the global mutation randomly selects one of the mutation operators compatible with the selected component and applies the mutation to it. 

In contrast, during \textit{local search}, the goal is to focus on one specific component of the Stateflow model. This is carried out when a mutation during global search has enhanced at least one of the three objective functions and has not deteriorated any of the other two objective functions. When this situation is given, the local mutation obtains the successful model, the mutated component, and the applied mutation operator (Line 18, Algorithm~\ref{alg:FlowRepair}). The local mutation only changes the component that has been enhanced. This is because it is likely that a change in a component that has enhanced at least one of the repair objectives has a high probability of being the buggy component. This theory aligns with the theory behind unified debugging~\cite{lou2020can,benton2021evaluating}. On the other hand, we follow a different strategy when selecting the mutation operator when compared to the global search. In this case, we give a 50\% probability that the same mutation operator will apply. For instance, if the relational operator replacement changes $>$ for $<$, and this has enhanced the verdict, we try to apply the same operator to see if another replacement (e.g., $<=$) is capable of repairing the patch. This decision was taken because we believe that, in many cases, not only the component to be mutated is important, but also the mutation operator. However, we also give a 50\% chance of applying another mutation operator, as many patches involve changing different parts (e.g., in a transition, you may change the relational operator but also a conditional operator).

\section{Empirical Evaluation} \label{sec:evaluation}
\label{sec:evaluation}

\subsection{Research Questions} \label{subsec:RQs}

In our evaluation, we aimed to answer the following research questions (RQs):

\begin{itemize}
    \item \textit{\textbf{RQ1 -- Repairability:} To what extent can \tool repair state-flow faults?}
    This RQ assesses the effectiveness of \tool to repair the Stateflow models in terms of generating plausible and valid patches.  
    \item \textit{\textbf{RQ2 -- Comparison with other algorithms:} How does \tool compare to a baseline algorithm?}
    We assess whether \tool is better than a (1+1)EA version of it, which is a common algorithm for APR in the context of other CPS studies.
    
\end{itemize}

\subsection{Dataset} \label{subsec:dataset}

Our dataset involves a total of 9 faulty models of 3 different case study systems. It is important to note that, while the number of faulty models is not large, the number of datasets involving real faults in Simulink models in general, and in Stateflow models in particular, is scarce. To the best of our knowledge, the only existing dataset involving faulty Stateflow models is provided by Ayesh et al.~\cite{ayesh2022two}, providing a total of 2 faulty models. We extend this dataset by including 7 new faulty models. These faulty models were extracted from students' projects where they had to develop a Stateflow controller. Note that the practice of using students' projects is common in other APR studies~\cite{gissurarson2022propr}. Although the number of bugs appears not to be large, it is important to note that (1) it is often difficult to experiment with a large set of faulty models in the context of CPSs due to their high computation~\cite{ValleSEIP2023,abdessalem2020automated}; and (2) this is the study of APR for CPSs with the largest number of bugs (i.e., 8 more bugs than Valle et al.~\cite{ValleSEIP2023}, 7 more bugs than Abdessalem et al.~\cite{abdessalem2020automated}, 5 more bugs than Jung et al.~\cite{jung2021swarmbug} and 3 more bugs than Lyu et al.~\cite{lyu2023autorepair}). We also considered to extend this dataset by including mutants. However, the mutation operators would be the same as the repair operators discussed in Section~\ref{sec:mutations}, thus, significant bias would be included in the evaluation, increasing the risk towards having flawed results. 

Therefore, in total, our dataset includes three different case study systems. The first one relates to two models of a pacemaker. Each model has one independent bug. Information about these buggy simulink models can be found on~\cite{ayesh2022two}. The second case study system involves the control of a fridge. We had three faulty models, but the second one had two faults in the model. Therefore, fridge\_1, fridge\_2 and fridge\_3 represent the buggy models. Fridge\_2a represents the second model with one of the bugs fixed and Fridge\_2b represents the second model with the other bug fixed. This would simulate how the APR tools behave when one of the bugs is manually fixed but the other one is not fixed yet. Lastly, the third case study system involves the model of an automated door that automatically opens and closes when detecting people in the building. The details (e.g., requirements) of these two case study systems can be found in the replication package.

As for the generated test cases, for each faulty model, we asked an independent developer to generate a failing and a passing test case. As for the oracles, we employed a regression oracle that provides the expected value of the system throughout the test execution. We opted for this kind of oracle because it is a widely adopted method by industrial CPS developers (e.g.,~\cite{sagardui2017configurable}). 

\subsection{Baseline Algorithm} \label{subsec:baseline}

We implemented a (1+1)EA version of our algorithm, which is in line with the algorithms used by Abdessalem et al.~\cite{abdessalem2020automated} and another CPS misconfiguration repair approach~\cite{ValleSEIP2023}. The algorithm is basically a version of our algorithm without the local search routine. It also maintains only a single partial archive, i.e., if a partial patch is found, the previous patch is removed from the archive. It is noteworthy, that the baseline algorithm employs the same strategy to select the component to be repaired as \tool. The algorithm is also stronger than the commonly employed Random Search baseline in other search-based studies.

\subsection{\tool Configuration} \label{subsec:toolconf}

\tool has two configurable parameters. Based on a set of preliminary experiments, we decided on the following values for each of them: (1) time budget: 1 hour; (2) the number of local tries: 30. The time budget can be increased in future studies depending on how much time the test execution takes for a specific CPS. As indicative for future replication studies, with this time budget and our case study systems, our algorithm generated between 150 and 1060 patches in each run, depending on the case study system.

As explained during the previous section, we also used the \textit{Tarantula} metric to obtain the suspiciousness score of our test cases, which is one of the most employed suspiciousness metrics~\cite{wong2016survey}.

\subsection{Evaluation Metrics}

For each run of the algorithm, we extract two metrics: (1) the number of plausible patches and (2) the number of valid patches. The former relates to the number of patches provided by \tool that passes the test suite, while the latter refers to the number of patches that are semantically equivalent to the patch proposed by the developer. 

\subsection{Experiment Runs}

Given the stochastic nature of search-based approaches, including \tool, we run the algorithm 5 times, both, for \tool and the baseline algorithm. As we employed 9 faulty models, and the given search budget was 1 hour, in total, the experiments lasted 90 hours (i.e., 5 (repetitions) $\times$ 9 (models) $\times$ 2 (algorithms) = 90 hours). We could not afford a higher number of repetitions, not only because of the long execution time of the algorithms, but also due to the significant manual effort for validating each of the plausible patches (i.e., with the current setup, we had to manually validate 686 plausible patches to check its semantic equivalence with the patch provided by the developer).

\subsection{Execution Environment}

We used MATLAB 2022b version for the models and \tool. The employed operating system was Windows 10, with 16GB RAM memory and an AMD Ryzen 7 5800HS processor with 8 cores and 16 threads.

\section{Analysis of the Results and Discussion} \label{sec:discussion}
\begin{figure*}[h]
\centering

\begin{subfigure}{0.3\textwidth}
%\centering
    \includegraphics[width=\textwidth, trim = 0 30 0 0]{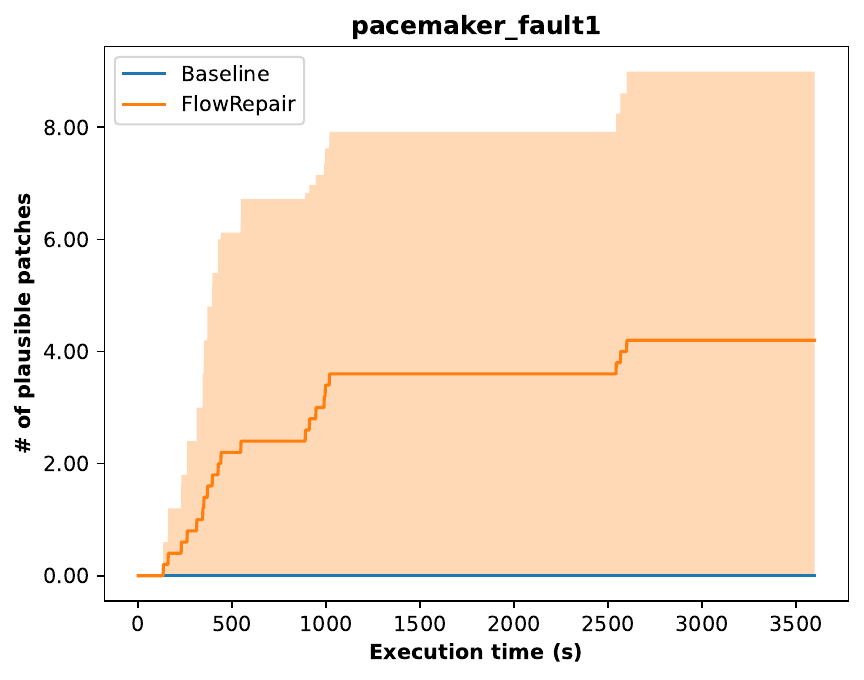}
    %\caption{c1}
    \label{fig:source}
\end{subfigure}
%\hspace*{\fill} % separation between the subfigures
\begin{subfigure}{0.302\textwidth}
%\centering
    \includegraphics[width=\textwidth, trim = 0 30 0 0] {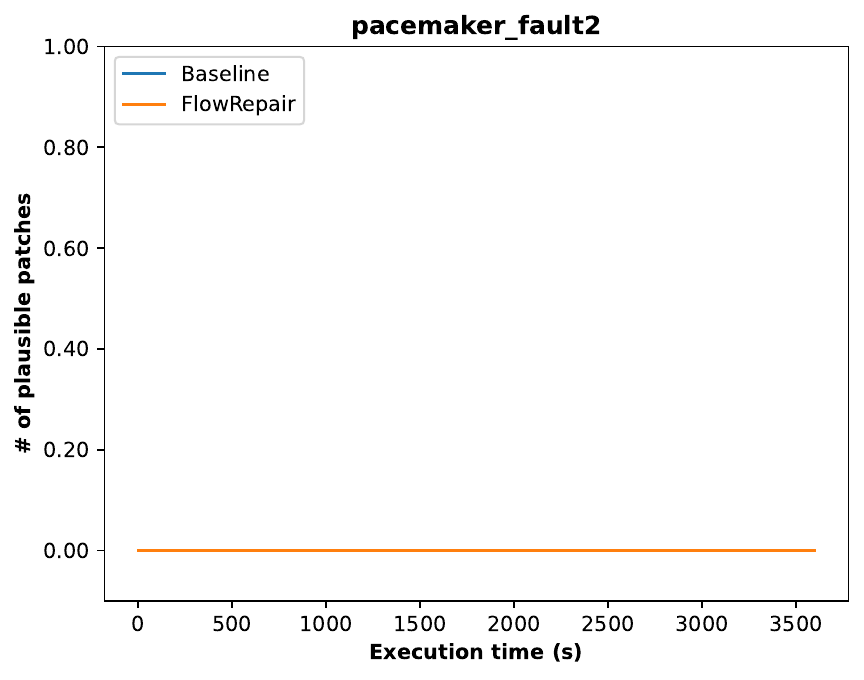} %{<left> <lower> <right> <upper>}
    %\caption{c2}
    \label{fig:mr1}
\end{subfigure}
\begin{subfigure}{0.305\textwidth}
%\centering
    \includegraphics[width=\textwidth, trim = 0 30 0 0] {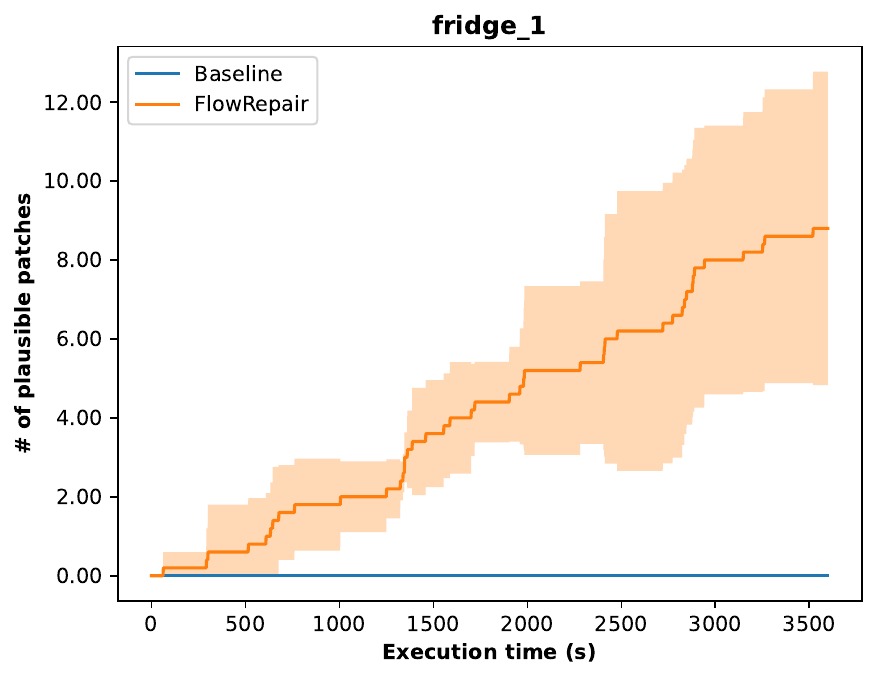} %{<left> <lower> <right> <upper>}
    %\caption{c2}
    \label{fig:mr1}
\end{subfigure}
\\
\begin{subfigure}{0.3\textwidth}
%\centering
    \includegraphics[width=\textwidth, trim = 0 30 0 0] {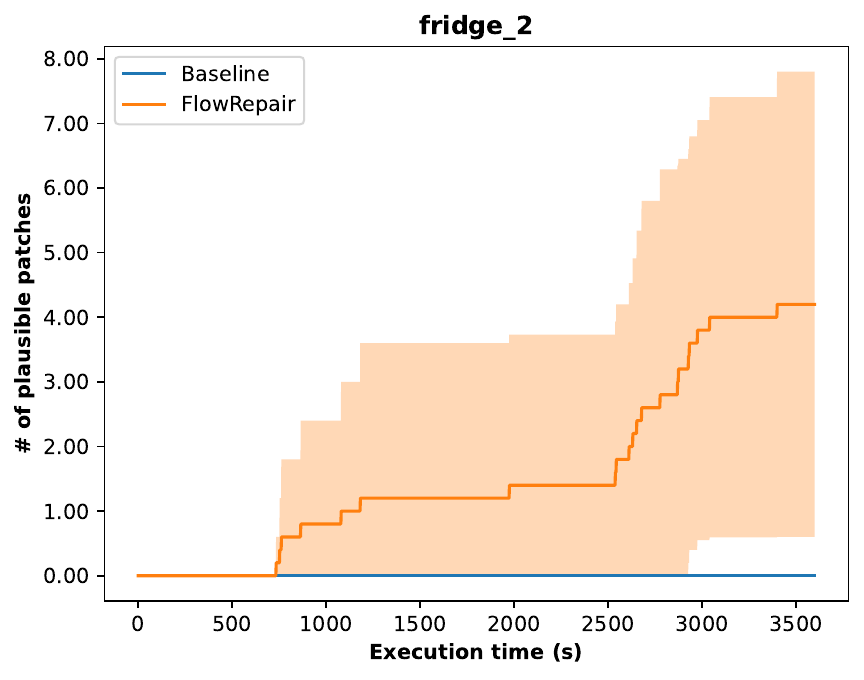} %{<left> <lower> <right> <upper>}
    %\caption{c2}
    \label{fig:mr1}
\end{subfigure}
\begin{subfigure}{0.305\textwidth}
%\centering
    \includegraphics[width=\textwidth, trim = 0 30 0 0] {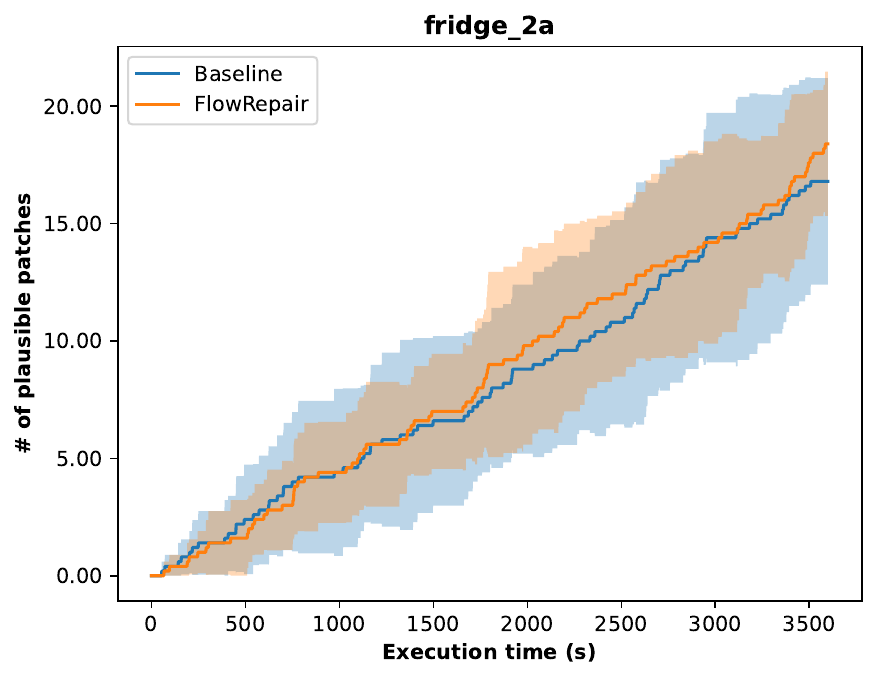} %{<left> <lower> <right> <upper>}
    %\caption{c2}
    \label{fig:mr1}
\end{subfigure}
\begin{subfigure}{0.305\textwidth}
%\centering
    \includegraphics[width=\textwidth, trim = 0 30 0 0] {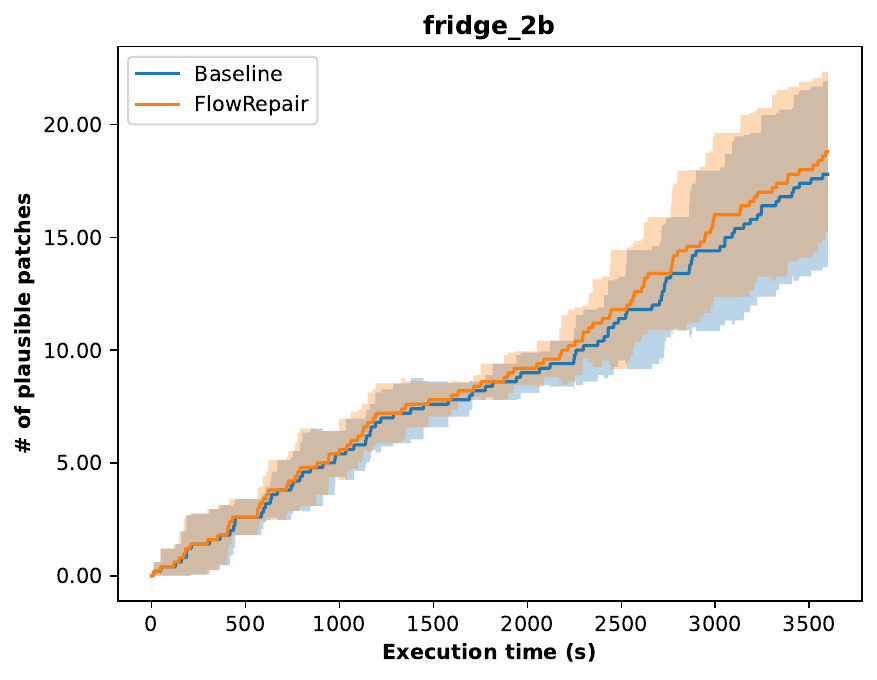} %{<left> <lower> <right> <upper>}
    %\caption{c2}
    \label{fig:mr1}
\end{subfigure}
\\
\begin{subfigure}{0.3\textwidth}
%\centering
    \includegraphics[width=\textwidth, trim = 0 30 0 0] {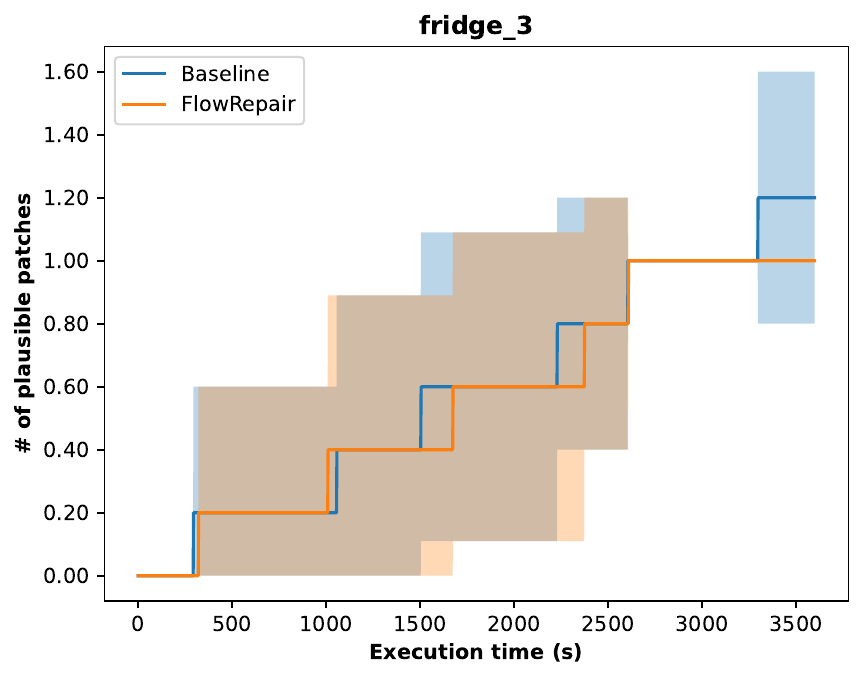} %{<left> <lower> <right> <upper>}
    %\caption{c2}
    \label{fig:mr1}
\end{subfigure}
\begin{subfigure}{0.305\textwidth}
%\centering
    \includegraphics[width=\textwidth, trim = 0 30 0 0] {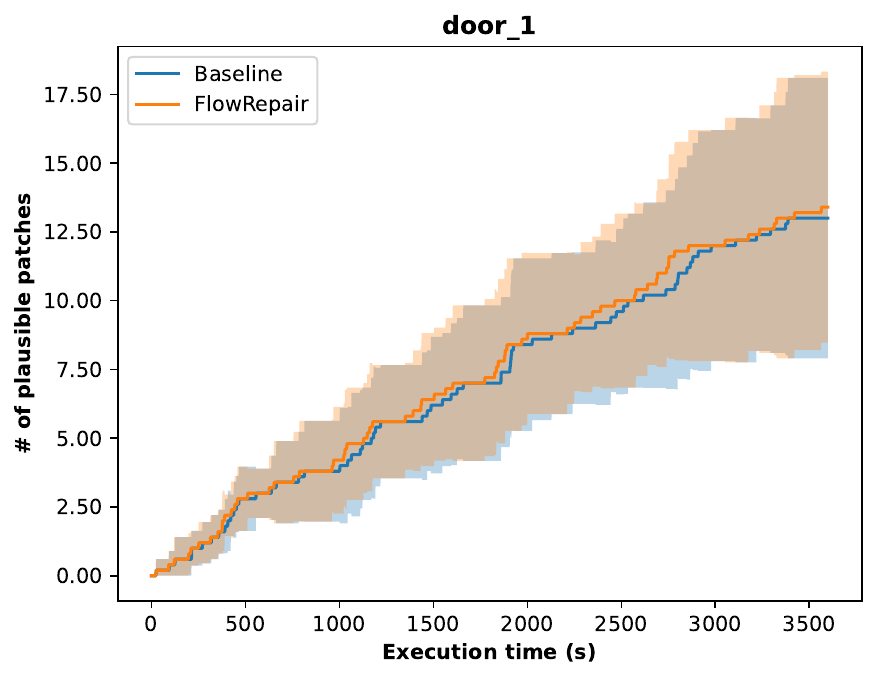} %{<left> <lower> <right> <upper>}
    %\caption{c2}
    \label{fig:mr1}
\end{subfigure}
\begin{subfigure}{0.305\textwidth}
%\centering
    \includegraphics[width=\textwidth, trim = 0 30 0 0] {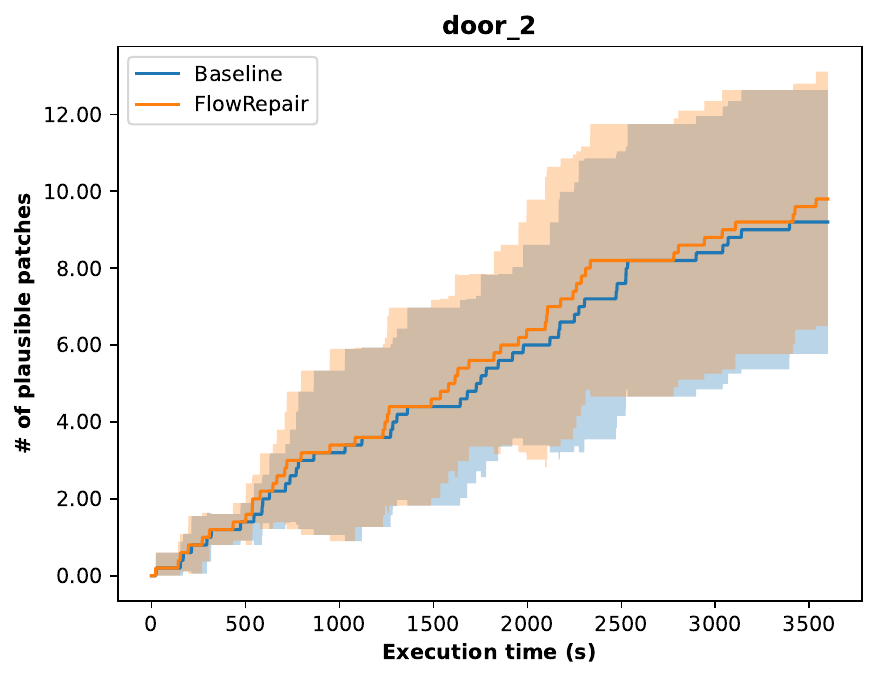} %{<left> <lower> <right> <upper>}
    %\caption{c2}
    \label{fig:mr1}
\end{subfigure}

\caption{Average number of plausible patches found by \tool and baseline over 5 different runs for one hour each}
\label{fig:results_noeffect}
\end{figure*}

\subsection{RQ1 -- Repairability of \tool}

Figure~\ref{fig:results_noeffect} depicts the obtained results in terms of the number of plausible patches found by \tool over time. Further, Table \ref{table:results} summarizes the global results of the entire algorithm executions. As can be seen, for all faulty models except for the pacemaker\_fault2, \tool was able to provide at least one plausible patch. It can also be appreciated from Figure~\ref{fig:results_noeffect} that the number of found plausible patches increased over time. In addition, it can be appreciated the standard deviation across the 5 runs. The example from fridge\_3 between 2600 and 3300 seconds shows no standard deviation since in all runs the number of plausible patches was the same (i.e., 1 plausible patch). The example from fridge\_2 is interesting, as this faulty model included two faults. Note, however, that we consider a plausible patch when the test suite is passed, i.e., both bugs need their corresponding fixes. In that example, it is seen that plausible patches were found after 750 seconds of execution, i.e., roughly after 12 minutes. As expected, more time was needed in this example because \tool focused on generating patches for one bug, and once this was fixed, focused on the second one. 

\begin{table*}[h]
\centering
\caption{Average, minimum and maximum values obtained for the 5 different runs. PP stands for Plausible Patch and VP stands for Valid Patch}
\label{table:results}
\centering
\resizebox{0.75\textwidth}{!}{
\begin{tabular}{lrrrrrrrrrrrr}
\toprule
 & \multicolumn{6}{c}{\textbf{\tool}} & \multicolumn{6}{c}{\textbf{Baseline}} \\ \cline{2-13} 
\multicolumn{1}{c}{\multirow{2}{*}{\textbf{Models}}} & \multicolumn{3}{c|}{\textbf{Plausible Patches}} & \multicolumn{3}{c|}{\textbf{Validated Patches}} & \multicolumn{3}{c|}{\textbf{Plausible Patches}} & \multicolumn{3}{c}{\textbf{Validated Patches}} \\ \cline{2-13} 
\multicolumn{1}{c}{} & \multicolumn{1}{c}{\textbf{mean}} & \multicolumn{1}{c}{\textbf{min}} & \multicolumn{1}{c|}{\textbf{max}} & \multicolumn{1}{c}{\textbf{mean}} & \multicolumn{1}{c}{\textbf{min}} & \multicolumn{1}{c|}{\textbf{max}} & \multicolumn{1}{c}{\textbf{mean}} & \multicolumn{1}{c}{\textbf{min}} & \multicolumn{1}{c|}{\textbf{max}} & \multicolumn{1}{c}{\textbf{mean}} & \multicolumn{1}{c}{\textbf{min}} & \multicolumn{1}{c}{\textbf{max}} \\ \hline
\textbf{Pacemaker\_1} & 4.2 & 0 & \multicolumn{1}{r|}{11} & 0.4 & 0 & \multicolumn{1}{r|}{1} & 0.0 & 0 & \multicolumn{1}{r|}{0} & 0.0 & 0 & 0 \\
\textbf{Pacemaker\_2} & 0.0 & 0 & \multicolumn{1}{r|}{0} & 0.0 & 0 & \multicolumn{1}{r|}{0} & 0.0 & 0 & \multicolumn{1}{r|}{0} & 0.0 & 0 & 0 \\
\textbf{Fridge\_1} & 8.8 & 3 & \multicolumn{1}{r|}{14} & 8.8 & 3 & \multicolumn{1}{r|}{14} & 0.0 & 0 & \multicolumn{1}{r|}{0} & 0.0 & 0 & 0 \\
\textbf{Fridge\_2} & 4.2 & 0 & \multicolumn{1}{r|}{8} & 0.4 & 0 & \multicolumn{1}{r|}{2} & 0.0 & 0 & \multicolumn{1}{r|}{0} & 0.0 & 0 & 0 \\
\textbf{Fridge\_2a} & 18.4 & 15 & \multicolumn{1}{r|}{24} & 13.4 & 9 & \multicolumn{1}{r|}{17} & 16.8 & 12 & \multicolumn{1}{r|}{25} & 13.4 & 10 & 17 \\
\textbf{Fridge\_2b} & 18.8 & 15 & \multicolumn{1}{r|}{25} & 7.4 & 4 & \multicolumn{1}{r|}{9} & 17.8 & 13 & \multicolumn{1}{r|}{25} & 6.6 & 4 & 9 \\
\textbf{Fridge\_3} & 1.0 & 1 & \multicolumn{1}{r|}{1} & 0.0 & 0 & \multicolumn{1}{r|}{0} & 1.2 & 1 & \multicolumn{1}{r|}{2} & 0.0 & 0 & 0 \\
\textbf{door\_1} & 13.4 & 8 & \multicolumn{1}{r|}{21} & 0.0 & 0 & \multicolumn{1}{r|}{0} & 13.0 & 8 & \multicolumn{1}{r|}{21} & 0.0 & 0 & 0 \\
\textbf{door\_2} & 9.8 & 6 & \multicolumn{1}{r|}{15} & 2.0 & 2 & \multicolumn{1}{r|}{6} & 9.0 & 5 & \multicolumn{1}{r|}{15} & 2.0 & 2 & 6 \\ \bottomrule
\end{tabular}
}
\end{table*}

\tool was not able to generate any plausible patch for pacemaker\_fault2 model. When having a closer look at potential reasons, we saw that the bug involved an incorrect value given to a variable. The bug involved the following assignation: ``\texttt{VENT\_CMP\_REF\\\_PWM = 125;}'', whereas the patch submitted by the developer was ``\texttt{VENT\_CMP\_REF\_PWM = VENT\_Sensitivity/5*100;}''. Given that the ``\texttt{VENT\_Sensitivity}'' variable was 3.5, the bug could eventually be fixed by applying the Numerical Replacement Operator mutation that changed the number 125 for 70, having this way a semantically equivalent fix to that proposed by the developer. Moreover, the tool could also apply several mutation operators to generate the exact same fix as the one proposed by the developer. However, although \tool had the capacity for generating such a fix, unfortunately, the buggy state in which this assignment was carried out involved a total of other 14 variable assignments. Therefore, even if a perfect fault location was assumed, the search space to fix this bug in such a model would be extremely large. That is, the way the model is developed restricts efficient APR. We believe that a better design of the Stateflow model could have helped repair the bug by \tool, although a more in-depth evaluation is required to confirm this hypothesis.

All these plausible patches were manually verified to see whether they were valid. For the five different runs, in six out of the eight buggy models \tool found at least a valid patch. In four of them, in all runs, there was at least one valid patch. We believe these results are quite competent, and, therefore, can summarize the first RQ as follows:

\begin{custombox}{RQ1}
\tool was able to find plausible patches for eight out of nine buggy models, and valid patches for six out of the nine buggy models. In four of the models, \tool consistently suggested valid patches.
\end{custombox}

\subsection{RQ2 -- Comparison with Baseline}

\tool was able to produce plausible patches for three out of nine  buggy models, (i.e., fridge\_1, fridge\_2 and pacemaker\_fault1) in which the selected baseline was not able to produce any plausible patch in any of the five runs. For the buggy model pacemaker\_fault1, \tool significantly surpassed the baseline algorithm, which found on average 0.4 plausible patches per run, whereas \tool generated, 4.2 plausible patches per run on average. We further confirmed there was statistical significance for this particular case by applying the Wilcoxon-rank sum test. When having a closer look at buggy models, we found that for many of the generated plausible patches, \tool made significant use of the local search routine. This might mean that to fix more complex bugs the strategy that combines global and local search makes the repair process to be significantly more efficient, something paramount in the context of CPSs. For the remaining buggy models in which plausible patches were found, when considering average values, \tool slightly surpassed the baseline algorithm in all buggy models except for Fridge\_3. However, the differences in all these cases were not statistically significant according to the Wilcoxon signed rank test we applied. In these buggy models, we found that, although for some models the patch was generated through the local search routine, in many other models, this was not necessary. Indeed, in a large portion of the models, the plausible patch was directly generated by applying a single edit on the initial buggy model. In those models, the fix could be provided by a pure random search version of \tool. We conjecture that the main reason is that the nature of such bugs is less complex than the ones in which the baseline algorithm failed to repair them.

Furthermore, the baseline was able to generate valid patches only in three of the buggy models, unlike \tool, which was able to generate valid patches in six of them. When comparing these three models, we found no statistical significance between \tool and the baseline.

\vspace{2pt}
\begin{custombox}{RQ2}
\tool is able to fix more bugs than the selected baseline. Given five tries, \tool found plausible patches in eight out of nine buggy models, whereas the baseline found plausible patches only in six models. \tool found valid patches in six buggy models, whereas the baseline found only in three of them.

\end{custombox}

\section{Threats to Validity} \label{sec:threats}
We now discuss which were the threats of the evaluation and how we tried to mitigate them:

\noindent\textbf{External Validity:} We applied \tool to only nine faulty models in three case studies, which may not be large enough to generalize our findings. These faults, however, were real faults and the only ones we found available for Stateflow models. The only dataset we found was from Ayesh et al.~\cite{ayesh2022two}. We complemented these with additional faulty models, extending it to nine faulty models. We note that in the context of CPSs, it is not common to have a large number of real faults available, as discussed in Section~\ref{subsec:dataset}. Nevertheless, we managed to have the largest dataset for APR in the context of CPSs.% which only had two faulty models, and we further extended it to nine models. Nonetheless, more faulty models are needed to generalize our results. 

\noindent\textbf{Internal Validity:} The configuration of various parameters of our algorithms is a concern that could affect the results. To mitigate this threat, we run preliminary experiments to find the optimal parameters of our algorithm, including the number of local tries. 

\noindent\textbf{Conclusion Validity:} We dealt with the randomness in our algorithm by repeating each algorithm run five times. Furthermore, to compare it with the baseline algorithm, we apply a relevant guide \cite{ArcuriGuide} to select the suitable statistical tests (i.e., the Wilcoxon signed-rank test).% \aitor{indicate}

\noindent\textbf{Construct Validity:} We chose two appropriate measures that are commonly used in other APR studies~\cite{jiang2021cure,lutellier2020coconut,fan2023automated,xia2023automated}, i.e., the number of plausible valid patches and the number of valid patches. Moreover, we used the same stopping criterion for our algorithm and the baseline algorithm for a fair comparison, i.e., time budget. %\aitor{@Pablo, can you find relevant APR papers (e.g., ICSE, ASE, ISSTA, FSE, TOSEM, TSE) in which these two metrics are used?}

\section{Related Work} \label{sec:relatedwork}
\label{sec:relatedwork}

Testing CPSs has been a widely investigated topic~\cite{CPSTestingSurvey}. This has been targeted from multiple perspectives, including test generation~\cite{menghi2020approximation,matinnejad2016automated,matinnejad2018test,riccio2020model,arrieta2017employing,fainekos2012verification,humeniuk2022search,nejati2019evaluating}, regression test optimization~\cite{arrieta2019pareto,arrieta2019search} and the test oracle problem~\cite{menghi2019generating,ayerdi2022performance,ayerdi2021generating}. In addition, testing CPSs has also been tackled focusing on different kinds of systems, e.g., autonomous vehicles~\cite{deepcollision,zhou2020deepbillboard,zhang2018deeproad,tian2018deeptest}, automated warehouses~\cite{uncertum,uncertest}, healthcare applications~\cite{sartaj2023hita}. We focus on Stateflow models, which is a notation for modeling the high-level control logic of Simulink models. Testing has been tackled from all these different perspectives in the context of Simulink models. For instance, Matinnejad et al.~\cite{matinnejad2016automated,matinnejad2018test} focused on test generation based on search algorithms that aim at increasing certain anti-patterns. Menghi et al.~\cite{menghi2020approximation} focused on test generation following a surrogate-assisted falsification-based approach. Nejati et al.~\cite{nejati2019evaluating} compared search-based techniques with model checking. Besides test generation, other existing approaches include oracle generation~\cite{menghi2019generating}, test system generation~\cite{arrieta2017automatic}, and test case prioritization~\cite{arrieta2019search}. However, all these approaches focus on testing. Conversely, our work focuses on repairing bugs specifically in the context of Stateflow models, which, to the best of our knowledge, has not been targeted yet.

Another line of research relates to debugging Simulink models. Some works focused on fault localization~\cite{liu2017improving,liu2016localizing,Liu2016,Liu2018,bartocci2022search}. Deshmukh et al.~\cite{deshmukh2018parameter} proposed a technique to localize misconfigurations. Other debugging works focused on other aspects, e.g., CPSDebug~\cite{CPSDebug} tests, debugs, and explains failures in Simulink/Stateflow models. Boufaied et al.~\cite{boufaied2023trace} focused on diagnosing CPS failures based on trace-checking input and output signals. Valle and Arrieta~\cite{valle2022towards} focused on reducing the failure-inducing test inputs of Simulink models. While all these studies focused on localizing bugs to ease the repairing process of buggy models, none of them targeted the automated repair task, which remains a manual activity in the context of Stateflow/Simulink models.

In the context of classical software systems, automated repair has been an intensive area of research aiming at reducing the laborious manual task of repairing software with an automated approach to generate patches for the software \cite{APRSurvey}. To this end, many approaches have been proposed applying different techniques, such as search algorithms~\cite{Arcuri2008,ASE2007,le2011genprog,yuan2018arja}, semantic analysis~\cite{nguyen2013semfix}, and machine-learning~\cite{Kim2013,DEAR,APRICSE2023,SobaniaChatGPT2023,tufano2018empirical,xia2023automated,xia2022less}. All these seminal works on APR are applied to general purpose software, but, as explained in the introduction, they cannot be directly applied to CPSs due to scalability issues, and would require, at least, several modifications. 

Despite this challenge, some works have focused on repairing CPSs. Abdessalem et al.~\cite{abdessalem2020automated} proposed Ariel, a tool that repairs interaction failures in automated driving systems. Jung et al.~\cite{jung2021swarmbug} proposed Swarmbug, a tool to repair misconfigurations of swarm aerial vehicles. Valle et al.~\cite{ValleSEIP2023} proposed an algorithm for repairing CPS misconfigurations applied to the context of elevator systems. Our works differ from these in two key perspectives. Firstly, these approaches do not focus on repairing the program itself, but either misconfigurations~\cite{ValleSEIP2023,jung2021swarmbug} or feature interactions~\cite{abdessalem2020automated}. Secondly, although they could eventually be adapted to other domains, these works focus on specific CPS domains and adapt their tools to target specific challenges from those domains, i.e., automotive~\cite{abdessalem2020automated}, robotics~\cite{jung2021swarmbug} and elevators~\cite{ValleSEIP2023}. Our work, in contrast, focuses on programs modeled in Simulink/Stateflow, which are common in many CPS domains~\cite{matinnejad2016automated,matinnejad2018test}, and does not make any assumptions nor implementation that is restricted to a specific system or domain.

To the best of our knowledge, similar to our approach, only two works target repair on Simulink models. Singh and Saha~\cite{SpecGuidedDebugging} focus on debugging, fault localization, and repairing Simulink models. However, their repair is restricted solely to parameters, unable to repair bugs due to, e.g., faults in relational, conditional, and mathematical operators, faults due to the incorrect destination of a transition, etc. Therefore, their approach would fail to repair the most common bugs in Simulink models (e.g., it would not be able to repair any of the faults of our benchmark). Another relevant work is AutoRepair~\cite{lyu2023autorepair}, which focuses on repairing neural networks embedded in CPSs. Specifically, AutoRepair isolates failure-inducing segments on a test case and uses a search algorithm to produce new data that satisfies an oracle. When this new data is found by the search algorithm, that data is used to extend the training data and retrain the neural network. Unlike AutoRepair, which focuses on repairing neural networks, \tool focuses on Stateflow models. AutoRepair and \tool are complementary: the former focuses on repairing bugs in neural networks, whereas the latter focuses on repairing Stateflow models.

\section{Conclusion and Future Work} \label{sec:conclusion}
\label{sec:conclusion}

Simulink is a well-known modeling and simulation tool employed in industry for the development of CPSs. As part of such modeling, Stateflow models are often used to model the control logic of CPSs. While significant effort has been devoted to advance on testing tasks of CPSs, the repair process has remained manual in such models. To this end, we presented \tool, an automated search-based tool to automatically repair Stateflow models. An empirical evaluation with 9 buggy models suggested that \tool is able to propose both valid and plausible patches in an affordable time budget for most buggy models, surpassing a baseline technique. However, among the plausible patches, we found that there were some over-fitting patches, which is a well-recognized problem in this context~\cite{goues2019automated,smith2015cure,yu2019alleviating}.%; (3) 

We identified the following research avenues that we will tackle in the near future. Firstly, we would like to analyze techniques to prioritize plausible patches for manual validation, reducing the time for finding valid patches. Secondly, we would also like to explore the potential of LLMs in proposing patches, for which we foresee applying different strategies (e.g., adding LLMs as mutators). Thirdly, we would like to explore and adapt techniques to alleviate the patch overfitting problem in the context of CPSs and Simulink models. Lastly, we would like to integrate this approach with other approaches (e.g., AutoRepair~\cite{lyu2023autorepair}), and propose other techniques to repair other Simulink parts (e.g., other blocks), offering a suite of APR techniques for Simulink models to industrial practitioners.

%Another of our findings suggests that some mutants are harder to detect than others. \aitor{WORK MORE THIS CONCLUSION}

\section*{Replication Package}

The tool and its evolution can be followed on GitHub: \url{https://github.com/aitorarrietamarcos/StateflowRepairTool}

The replication package and specific version used in our evaluation is available on Zenodo: \url{https://zenodo.org/records/10936238}

\section*{Acknowledgments}
Aitor Arrieta and Pablo Valle are part of the Software and Systems Engineering research group of Mondragon Unibertsitatea (IT1519-22), supported by the Department of Education, Universities and Research of the Basque Country. Shaukat Ali is also supported by the Co-evolver project (Project \#286898), funded by the Research Council of Norway.  
%Bibliography
\bibliographystyle{unsrt}  
\bibliography{references}

\end{document}